\documentstyle[prl,aps]{revtex}

\begin{document}
\draft
\title{Local and Nonlocal Properties of Werner States}
\author{Tohya Hiroshima\cite{NEC1} and Satoshi Ishizaka\cite{NEC2}}
\address{Fundamental Research Laboratories,\\
NEC Corporation, 34 Miyukigaoka, Tsukuba 305-8501, Japan}
\maketitle

\begin{abstract}
We consider a special kind of mixed states -- a {\it Werner derivative},
which is the state transformed by nonlocal unitary -- local or nonlocal --
operations from a Werner state. We show the followings. (i) The amount of
entanglement of Werner derivatives cannot exceed that of the original Werner
state. (ii) Although it is generally possible to increase the entanglement
of a single copy of a Werner derivative by LQCC, the maximal possible
entanglement cannot exceed the entanglement of the original Werner state.
The extractable entanglement of Werner derivatives is limited by the
entanglement of the original Werner state.
\end{abstract}


Quantum entanglement plays an essential role in various types of quantum
information processing, including quantum teleportation \cite{BBCJPW},
superdense coding \cite{BW}, quantum cryptographic key distribution \cite
{Ekert}, and quantum computation \cite{EJVP}. Since the best performance of
such tasks requires maximally entangled states (Bell singlet states), one of
the most important entanglement manipulations is the entanglement
purification or distillation \cite{Gisin,HHH,BBPSSW,BDSW}, namely, the
process extracting maximally entangled states from input states. Most of
protocols for entanglement purification (distillation) proposed so far \cite
{BBPSSW,BDSW} utilize the collective operations on many copies of a given
state $\rho $. Strictly speaking, these protocols rely only on the
properties of $\rho ^{\otimes N}$ with large $N$ and have no direct
relevance to the intrinsic properties of the individual state $\rho $. In
fact, it has been shown that there exist no purification protocols utilizing
local quantum operations and classical communications (LQCC) producing a
pure singlet from a {\it single} copy of a given mixed state of two qubits 
\cite{Kent}. If we are not available to many copies of a given mixed state
but a single one, the only task we can do by LQCC is to enhance the amount
of entanglement to some extent. However, there exist entangled mixed state,
for which even such a restricted task is also not successful \cite
{Kent,LMP,KLM}. Therefore, it is of fundamental importance to clarify the
limit of entanglement manipulations of a single copy of a given mixed state
for deeper understanding of the nature of mixed state entanglement.

We consider in this paper a special kind of mixed states -- a {\it Werner
derivative}, which refers to the state transformed by unitary -- local or
nonlocal -- operations from a Werner state \cite{Werner}. We show the
followings. (i) The amount of entanglement of Werner derivatives cannot
exceed that of the original Werner state. (ii) Although it is generally
possible to increase the entanglement of a single copy of a Werner
derivative by LQCC, the maximal possible entanglement cannot exceed the
entanglement of the original Werner density matrix. The extractable
entanglement of Werner derivatives is limited by the entanglement of the
original Werner state. Here, the extractable entanglement of a given state $%
\rho $ is referred to as the maximal possible entanglement obtained by LQCC
applied to a single copy of $\rho $ ({\it single-state LQCC}) \cite{KLM}.
The first point (i) is the direct consequence of the results presented in
our recent work \cite{IH}, that is, a Werner state belongs to a set of {\it %
maximally entangled mixed states}, in which the amount of entanglement
cannot be increased by applying any unitary operations. The second point
(ii) is our main result.

The degree of entanglement of mixed states of two qubits is customarily
measured by the entanglement of formation (EOF) \cite{BDSW}. The EOF for a
two-party pure state is defined as the von Neumann entropy of the reduced
density matrix associated with one of the parties. The EOF of a bipartite
mixed state is defined as $E_{F}(\rho )=\min \sum_{i}p_{i}E_{F}\left( \left|
\psi_{i}\right\rangle \left\langle \psi _{i}\right| \right) $, where the
minimum is taken over all possible decomposition of $\rho $ into pure
states, $\rho=\sum_{i}p_{i}\left| \psi_{i}\right\rangle \left\langle \psi
_{i}\right| $. In $2\times 2$ systems the closed form for EOF is known \cite
{Wooters}; 
\begin{equation}
E_{F}(\rho )=H\left( \frac{1+\sqrt{1-C^{2}}}{2}\right),
\end{equation}
with $H(x)=-x\log _{2}x-(1-x)\log _{2}(1-x)$. The nonnegative real number $%
C=\max \{0,\lambda _{1}-\lambda _{2}-\lambda _{3}-\lambda _{4}\}$ is called
a concurrence, where $\lambda _{i}$ are the square roots of eigenvalues of
positive matrix $\rho \widetilde{\rho }$ in descending order. The
spin-flipped density matrix $\widetilde{\rho }$ is defined as $\widetilde{%
\rho } =\sigma _{2}\otimes \sigma _{2}\rho ^{*}\sigma _{2}\otimes \sigma
_{2} $, where asterisk denotes complex conjugation in the standard basis $%
\left\{\left| 00\right\rangle ,\left| 01\right\rangle ,\left|
10\right\rangle ,\left| 11\right\rangle \right\} $ and $\sigma _{i}$, $%
i=1,2,3$, are usual Pauli matrices. Since $E_{F}$ is a monotonic function of 
$C$ and $C$ ranges from zero to one, the concurrence $C$ is also a measure
of entanglement.

Before verifying our main result, we firstly show that a Werner state
belongs to a family of maximally entangled mixed states. Although the
argument based on the convexity of concurrence is presented in Ref. \cite{IH}%
, we follows here the direct calculations for later convenience. A Werner
state in $2\times 2$ systems takes the following form, 
\begin{equation}
\rho _{W}=\frac{1-F}{3}{\bf I}_{4}+\frac{4F-1}{3}\left| \Psi
^{-}\right\rangle \left\langle \Psi ^{-}\right| ,
\end{equation}
where ${\bf I}_{n}$ denotes the $n\times n$ identity matrix and $\left| \Psi
^{-}\right\rangle =\left( \left| 01\right\rangle -\left| 10\right\rangle
\right) /\sqrt{2}$ the singlet state. The Werner state $\rho _{W}$ is
characterized by a single real parameter $F$ called fidelity. This quantity
measures the overlap of the Werner state with a Bell state. The concurrence
of $\rho _{W}$ is simply given by $C(\rho _{W})=\max \{0,2F-1\}$; for $F\leq
1/2$ the Werner state is unentangled, while for $1/2<F\leq 1$ it is
entangled. We assume $1/2<F\leq 1$ so that $C(\rho _{W})=2F-1$ in the
following.

The nonlocal unitary transformation, $U\in U(4)$, brings $\rho _{W}$ to a
new density matrix of the form, 
\begin{equation}
\rho =\frac{1-F}{3}{\bf I}_{4}+\frac{4F-1}{3}\left| \psi \right\rangle
\left\langle \psi \right|,  \label{eq:WD}
\end{equation}
where $\left| \psi \right\rangle =U\left| \Psi ^{-}\right\rangle $. Because $%
U$ preserves the rank of states, $\left| \psi \right\rangle $ is still a
pure (rank of one) state vector but is generally less entangled and it can
be written in a Schmidt decomposed form, $\left| \psi \right\rangle =\sqrt{a}%
\left|00\right\rangle +\sqrt{1-a}\left| 11\right\rangle $ with $1/2\leq
a\leq 1$. The nonlocal unitary transformation $U$ is thus parametrized by a
single real number $a$. The Peres-Horodecki criterion (the partial
transposition test) \cite{Peres,Horodecki} tells us that the Werner
derivative described by Eq. (\ref{eq:WD}) is entangled if and only if 
\begin{equation}
\frac{1}{2}\leq a<\frac{1}{2}\left( 1+\frac{\sqrt{3(4F^{2}-1)}}{4F-1}\right).
\end{equation}
The range of parameter $a$ is assumed to be limited by above inequalities so
that $\rho $ is always entangled. The square roots of eigenvalues of $\rho 
\widetilde{\rho }$ are calculated as 
\begin{equation}
\lambda _{1}=\frac{(4F-1)G_{+}}{3},  \label{eq:S1}
\end{equation}
\begin{equation}
\lambda _{2}=\frac{(4F-1)G_{-}}{3},  \label{eq:S2}
\end{equation}
and 
\begin{equation}
\lambda _{3}=\lambda _{4}=\frac{1-F}{3},  \label{eq:S3}
\end{equation}
which are sorted in decreasing order. In Eqs. (\ref{eq:S1}) and (\ref{eq:S2}%
), 
\begin{equation}
G_{\pm }=\left[ 2a(1-a)+G\pm 2\sqrt{a(1-a)(a(1-a)+G)}\right] ^{\frac{1}{2}},
\end{equation}
with $G=3F(1-F)/(4F-1)^{2}$. The concurrence of $\rho $, $C(\rho
)=\lambda_{1}-\lambda_{2}-\lambda _{3}-\lambda _{4}$, is given by 
\begin{equation}
C(\rho )=\frac{4F-1}{3}\left( G_{+}-G_{-}\right) -\frac{2}{3}(1-F).
\end{equation}
The problem is to find the maximal value of $C(\rho )$. We have that 
\begin{equation}
\frac{d}{da}C(\rho )=\frac{1}{6}\frac{(4F-1)(1-2a)}{\sqrt{a(1-a)(a(1-a)+G)}}%
\left( G_{+}+G_{-}\right),
\end{equation}
which is clearly nonpositive for $a\geq 1/2$. It follows that the maximal $%
C(\rho )$ is achieved only for $a=1/2$. The maximal value of the concurrence
is calculated as $2F-1$. Therefore, the EOF of Werner derivatives $%
E_{F}(\rho )$ cannot exceed the EOF of the original Werner state $%
E_{F}(\rho_{W})$; a Werner state is indeed a member of a set of maximally
entangled mixed states.

Now let us turn to the proof of our main result. The Werner derivative $\rho 
$ given by Eq. (\ref{eq:WD}) can be also written as 
\begin{equation}
\rho =\frac{1}{4}{\bf I}_{4}+\frac{4F-1}{12}\left[ (2a-1)\left( {\bf I}%
_{2}\otimes \sigma _{3}+\sigma _{3}\otimes {\bf I}_{2}\right) +2\sqrt{a(1-a)}%
\left( \sigma _{1}\otimes \sigma _{1}+\sigma _{2}\otimes \sigma _{2}\right)
+\sigma _{3}\otimes \sigma _{3}\right] .
\end{equation}
Since the coefficient vectors of
${\bf I}_{2}\otimes \mbox{\boldmath $\sigma $} $ or
$ \mbox{\boldmath $\sigma $} \otimes {\bf I}_{2}$ are nonzero 
$ \left[ \mbox{\boldmath $\sigma $}
=\left( \sigma_{1},\sigma _{2},\sigma _{3}\right)
 \right] $,
it is possible to increase
the EOF of $\rho $ by a single-state LQCC \cite{KLM}. As shown below,
however, the maximum EOF thus obtained is still less than or equal to the
EOF of the original Werner state. According to Theorem 3 in Ref. \cite{KLM},
there exist a single-state LQCC mapping $\rho $ to a Bell diagonal state $%
\rho ^{\prime }$ with maximal possible EOF of the form, 
\begin{equation}
\rho ^{\prime }=\frac{1}{4}\left( {\bf I}_{4}+\sum_{i=1}^{3}r_{i}\sigma
_{i}\otimes \sigma _{i}\right) ,
\end{equation}
with $r_{1}\leq r_{2}\leq r_{3}\leq 0$. The square roots of eigenvalues of $%
\rho ^{\prime }\widetilde{\rho ^{\prime }}$ in descending order are $\lambda
_{1}^{\prime }=(1-r_{1}-r_{2}-r_{3})/4$, $\lambda _{2}^{\prime
}=(1-r_{1}+r_{2}+r_{3})/4$, $\lambda _{3}^{\prime }=(1+r_{1}-r_{2}+r_{3})/4$%
, and $\lambda _{4}^{\prime }=(1+r_{1}+r_{2}-r_{3})/4$. Since the ratio $%
\lambda _{i}^{\prime }/\lambda _{j}^{\prime }$ are invariant under LQCC, $%
\lambda _{i}^{\prime }/\lambda _{4}^{\prime }=\lambda _{i}/\lambda _{4}$ $%
(i=1,2,3)$, where $\lambda _{i}$ are given by Eqs. (\ref{eq:S1}), (\ref
{eq:S2}), and (\ref{eq:S3}). Therefore, the concurrence of $\rho ^{\prime }$%
, $C(\rho ^{\prime })=\lambda _{1}^{\prime }-\lambda _{2}^{\prime }-\lambda
_{3}^{\prime }-\lambda _{4}^{\prime }$, can be expressed in terms of $%
\lambda _{i}$ as follows, 
\begin{equation}
C(\rho ^{\prime })=\frac{\lambda _{1}-\lambda _{2}-\lambda _{3}-\lambda _{4}%
}{\lambda _{1}+\lambda _{2}+\lambda _{3}+\lambda _{4}}.  \label{eq:XE}
\end{equation}
Inserting the explicit forms of $\lambda _{i}$ into this equation, we obtain 
\begin{equation}
C(\rho ^{\prime })-C(\rho _{W})=2\frac{(1-F)G_{+}-FG_{-}-2(1-F)/(4F-1)}{%
G_{+}+G_{-}+2(1-F)/(4F-1)}.
\end{equation}
The denominator of the right hand side of this equation is strictly positive
so that it suffices to verify the numerator is less than or equal to zero in
order to show $C(\rho ^{\prime })\leq C(\rho _{W})$. We have that 
\begin{equation}
\frac{d}{da}\left[ (1-F)G_{+}-FG_{-}\right] =\frac{1}{2}\frac{1-2a}{\sqrt{%
a(1-a)(a(1-a)+G)}}\left[ (1-F)G_{+}+FG_{-}\right] ,  \label{eq:DF}
\end{equation}
which is clearly nonpositive for $a\geq 1/2$. It follows that maximal value
of $\left[ (1-F)G_{+}-FG_{-}\right] $ is achieved for $a=1/2$ and it turns
out to be $2(1-F)/(4F-1)$. Therefore, numerator in the right hand side of
Eq. (\ref{eq:DF}) is strictly less than or equal to zero. Hence $C(\rho
^{\prime })\leq C(\rho _{W})$ so that $E_{F}(\rho ^{\prime })\leq E_{F}(\rho
_{W})$. It should be noted that unitary transformation with $a=1/2$ is just
a local unitary transformation; $\left| 0\right\rangle _{A}\rightarrow
\left| 0\right\rangle _{A}$, $\left| 1\right\rangle _{A}\rightarrow \left|
1\right\rangle _{A}$, $\left| 0\right\rangle _{A}\rightarrow -\left|
1\right\rangle _{A}$, and $\left| 1\right\rangle _{A}\rightarrow \left|
0\right\rangle _{A}$ such that $\left| \Psi ^{-}\right\rangle _{AB}=\left(
\left| 01\right\rangle _{AB}-\left| 10\right\rangle _{AB}\right) /\sqrt{2}%
\rightarrow $ $\left( \left| 00\right\rangle _{AB}+\left| 11\right\rangle
_{AB}\right) /\sqrt{2}$. The state $\rho $ is, therefore, equivalent to $%
\rho _{W}$ up to local unitary transformations and the present result is
reduced to that of Ref. \cite{LMP}. It implies the following. If we bring a
Werner state $\rho _{W}$ to one of Werner derivatives $\rho $ by {\it %
essentially nonlocal} unitary transformations, the extractable entanglement
of $\rho $ is strictly below the EOF of the original Werner state. Our main
result can be also stated in other words that the EOF of a Werner state
cannot be increased by a single-state LQCC followed by nonlocal unitary
transformations. This property is unique to Werner states, as shown below.
If another state $\rho $ which does not belong to a family of Werner states
has the property stated above, it must be one of the maximally entangled
mixed states; otherwise $E_{F}(\rho )$ could be increased by nonlocal
unitary transformation. The maximally entangled mixed states take the
following form \cite{IH}, 
\begin{equation}
\rho =p_{1}\left| \Psi ^{-}\right\rangle \left\langle \Psi ^{-}\right|
+p_{2}\left| 00\right\rangle \left\langle 00\right| +p_{3}\left| \Psi
^{+}\right\rangle \left\langle \Psi ^{+}\right| +p_{4}\left| 11\right\rangle
\left\langle 11\right| ,
\end{equation}
where $\left| \Psi ^{+}\right\rangle =\left( \left| 01\right\rangle +\left|
10\right\rangle \right) /\sqrt{2}$ and $p_{i}$ are eigenvalues of $\rho $ in
decreasing order $(p_{1}\geq p_{2}\geq p_{3}\geq p_{4}\geq 0)$. The state $%
\rho $ can be also written as 
\begin{equation}
\rho =\frac{1}{4}\left[ {\bf I}_{4}+(p_{2}-p_{4})\left( {\bf I}_{2}\otimes
\sigma _{3}+\sigma _{3}\otimes {\bf I}_{2}\right) -(p_{1}-p_{3})\left(
\sigma _{1}\otimes \sigma _{1}+\sigma _{2}\otimes \sigma _{2}\right)
-(p_{1}-p_{2}+p_{3}-p_{4})\sigma _{3}\otimes \sigma _{3}\right] .
\end{equation}
If $p_{2}\neq p_{4}$, the coefficient vectors of
${\bf I}_{2}\otimes \mbox{\boldmath $\sigma $} $ or
$ \mbox{\boldmath $\sigma $} \otimes {\bf I}_{2}$ are nonzero 
and the EOF of $%
\rho $ can be increased further by a single-state LQCC, which contradicts
the assumed property of $\rho $. Therefore, the equality $p_{2}=p_{4}$ must
hold, which implies $p_{2}=p_{3}=p_{4}=(1-p_{1})/3$. It follows that the
state $\rho $ takes the form, 
\begin{equation}
\rho =\frac{1-p_{1}}{3}{\bf I}_{4}+\frac{4p_{1}-1}{3}\left| \Psi
^{-}\right\rangle \left\langle \Psi ^{-}\right|.
\end{equation}
Hence, $\rho $ must be a Werner state.

Finally, we mention that Eq. (\ref{eq:XE}) gives the general expression for
the extractable entanglement of a given entangled state $\rho $ of two
qubits. It has the form of the concurrence of $\rho $, $C(\rho )=\lambda
_{1}-\lambda _{2}-\lambda _{3}-\lambda _{4}$, modified by an enhancement
factor $(\lambda _{1}+\lambda _{2}+\lambda _{3}+\lambda _{4})^{-1}$. For
Bell diagonal states, including Werner states, $\rho =\widetilde{\rho }$ so
that the square roots of eigenvalues of $\rho \widetilde{\rho }$ are same as
the eigenvalues of $\rho $. Therefore, $\lambda _{1}+\lambda _{2}+\lambda
_{3}+\lambda _{4}=1$ and the enhancement factor is one. It follows directly
that we cannot extract higher EOF from a Bell diagonal state by a
single-state LQCC as argued in Ref. \cite{KLM}. For pure states of rank one, 
$C(\rho ^{\prime })=1$, which indicates that it is always possible to
extract a Bell singlet state as expected.

In summary, combined the present results with previously obtained ones \cite
{LMP,IH}, the following peculiar property of a Werner state of two qubits
has been revealed; its EOF cannot be increased (i) by LQCC, (ii) by nonlocal
unitary transformations, and (iii) by LQCC followed by nonlocal unitary
transformations. We hope that our results presented in this paper would lead
to a proper classification of entangled mixed states.


\begin{references}
\bibitem[*]{NEC1}  E-mail address: tohya@frl.cl.nec.co.jp
\bibitem[**]{NEC2}  E-mail address: isizaka@frl.cl.nec.co.jp

\bibitem{BBCJPW}  C. H. Bennett, G. Brassard, C. Crepeau, R. Jozsa, A. Peres, 
and W. K. Wootters, Phys. Rev. Lett. {\bf 70}, 1895 (1993).
\bibitem{BW}  C. H. Bennett and S. J. Wiesner, Phys. Rev. Lett. {\bf 69},
2881 (1993).
\bibitem{Ekert}  A. Ekert, Phys. Rev. Lett. {\bf 67}, 661 (1991).
\bibitem{EJVP}  For reviews, see A. Ekert and R. Jozsa, Rev. Mod. Phys. {\bf %
68}, 733 (1996); V. Vedral and M. B. Plenio, Prog. Quantum Electron. {\bf 22}%
, 1 (1998).
\bibitem{Gisin} N. Gisin, Phys. Lett. A {\bf 210}, 151 (1996).
\bibitem{HHH} M. Horodecki, P. Horodecki, and R. Horodecki, Phys. Rev. Lett. {\bf 78}, 574 (1997).
\bibitem{BBPSSW}  C. H. Bennett, G. Brassard, S. Popescu, B. Schumacher, J.
Smolin, and W. K. Wootters, Phys. Rev. Lett. {\bf 76}, 722 (1996).
\bibitem{BDSW}  C. H. Bennett, D. P. Di Vincenzo, J. Smolin, and W. K.
Wootters, Phys. Rev. A {\bf 54}, 3824 (1996).
\bibitem{Kent}  A. Kent, Phys. Rev. Lett. {\bf 81}, 2839 (1998).
\bibitem{LMP}  N. Linden, S. Massar, and S. Popescu, Phys. Rev. Lett. {\bf 81%
}, 3279 (1998).
\bibitem{KLM}  A. Kent, N. Linden, and S. Massar, Phys. Rev. Lett. {\bf 83},
2656 (1999).
\bibitem{Werner}  R. Werner, Phys. Rev. A {\bf 40}, 4277 (1989).
\bibitem{IH}  S. Ishizaka and T. Hiroshima, quant-ph/0003023.
\bibitem{Wooters}  W. K. Wooters, Phys. Rev. Lett. {\bf 80}, 2245 (1998).
\bibitem{Peres}  A. Peres, Phys. Rev. Lett. {\bf 77}, 1413 (1996).
\bibitem{Horodecki}  P. Horodecki, Phys. Lett. A, {\bf 232}, 333 (1997).
\end{references}
\end{document}